\documentstyle[pre,aps,epsfig]{revtex} 

\begin{document}
\title{Relaxation kinetics of biological dimer adsorption models}

\author{A. Vilfan}
\address{Cavendish
    Laboratory, Madingley Road, Cambridge CB3 0HE, UK}
\author{E. Frey}
\address{Abteilung Theorie, Hahn-Meitner-Institut, Glienickerstr. 100, 
D-14109 Berlin}
\author{F. Schwabl}
\address{Institut f\"ur Theoretische Physik, Technische
    Universit\"at M\"unchen, D--85747 Garching}

\date{17/8/2001}

\maketitle

\begin{abstract}
  We discuss the relaxation kinetics of a one-dimensional dimer
  adsorption model as recently proposed for the binding of biological
  dimers like kinesin on microtubules. The non-equilibrium dynamics
  shows several regimes: irreversible adsorption on short time scales,
  an intermediate plateau followed by a power-law regime and finally
  exponential relaxation towards equilibrium.  In all four regimes we
  give analytical solutions.  The algebraic decay and the scaling
  behaviour can be explained by mapping onto a simple
  reaction-diffusion model. We show that there are several
  possibilities to define the autocorrelation function and that they
  all asymptotically show exponential decay, however with different
  time constants.  Our findings remain valid if there is an attractive
  interaction between bound dimers.
\end{abstract}

\begin{figure}
\begin{tabular}[b]{ll}
{\small Fig.~\ref{fig_reactions}a.}~~
{\epsfxsize=0.25\textwidth \epsffile{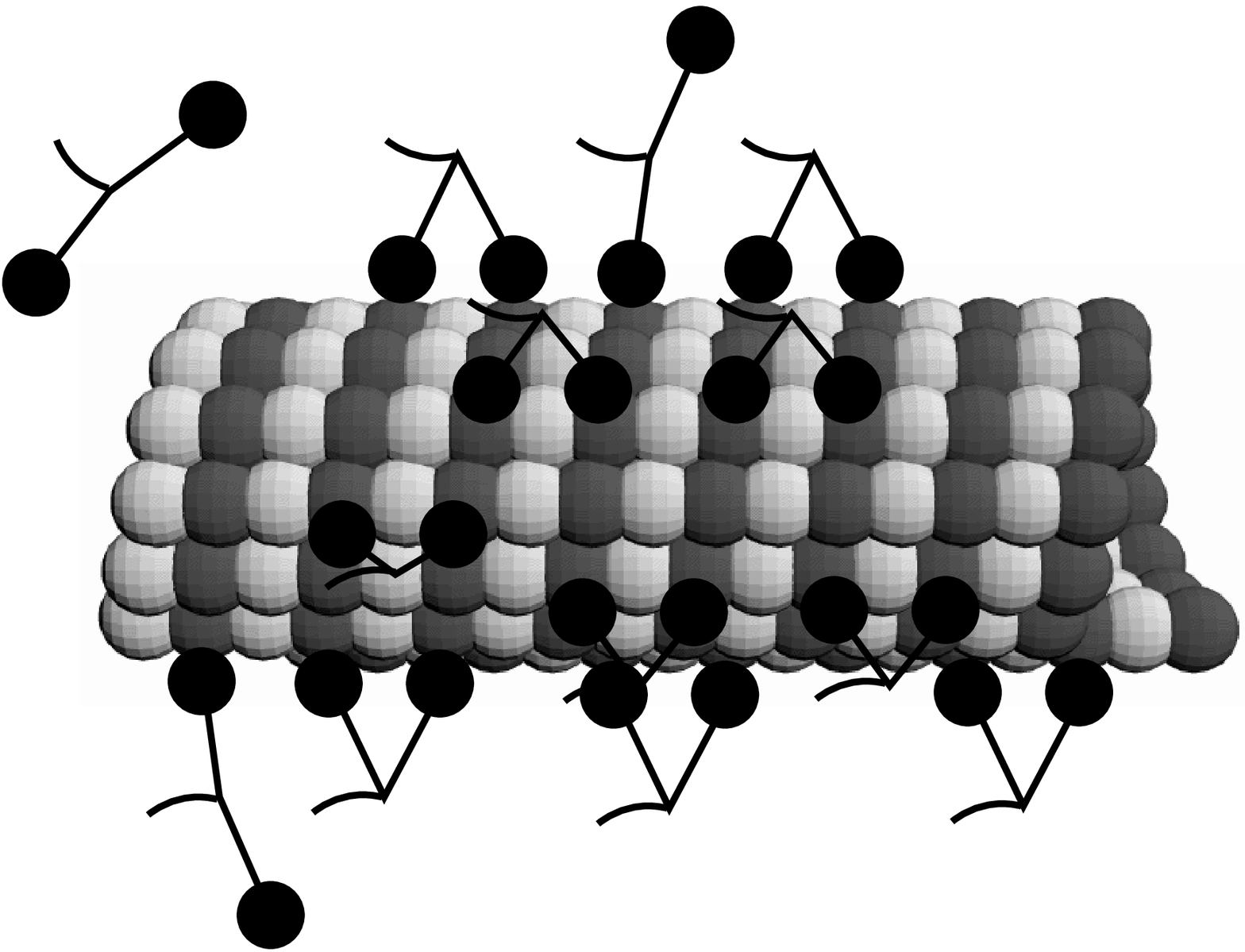}}\\
{\epsfxsize=0.45\textwidth \epsffile{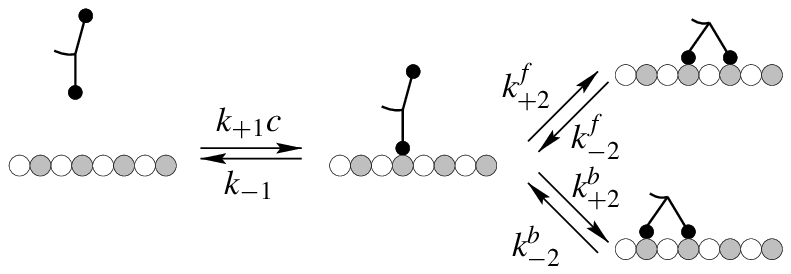}}\\
{\small Fig.~\ref{fig_reactions}b.}
\end{tabular}
\nolinebreak
\begin{tabular}[b]{ll}
{\epsfxsize=0.47\textwidth\epsffile{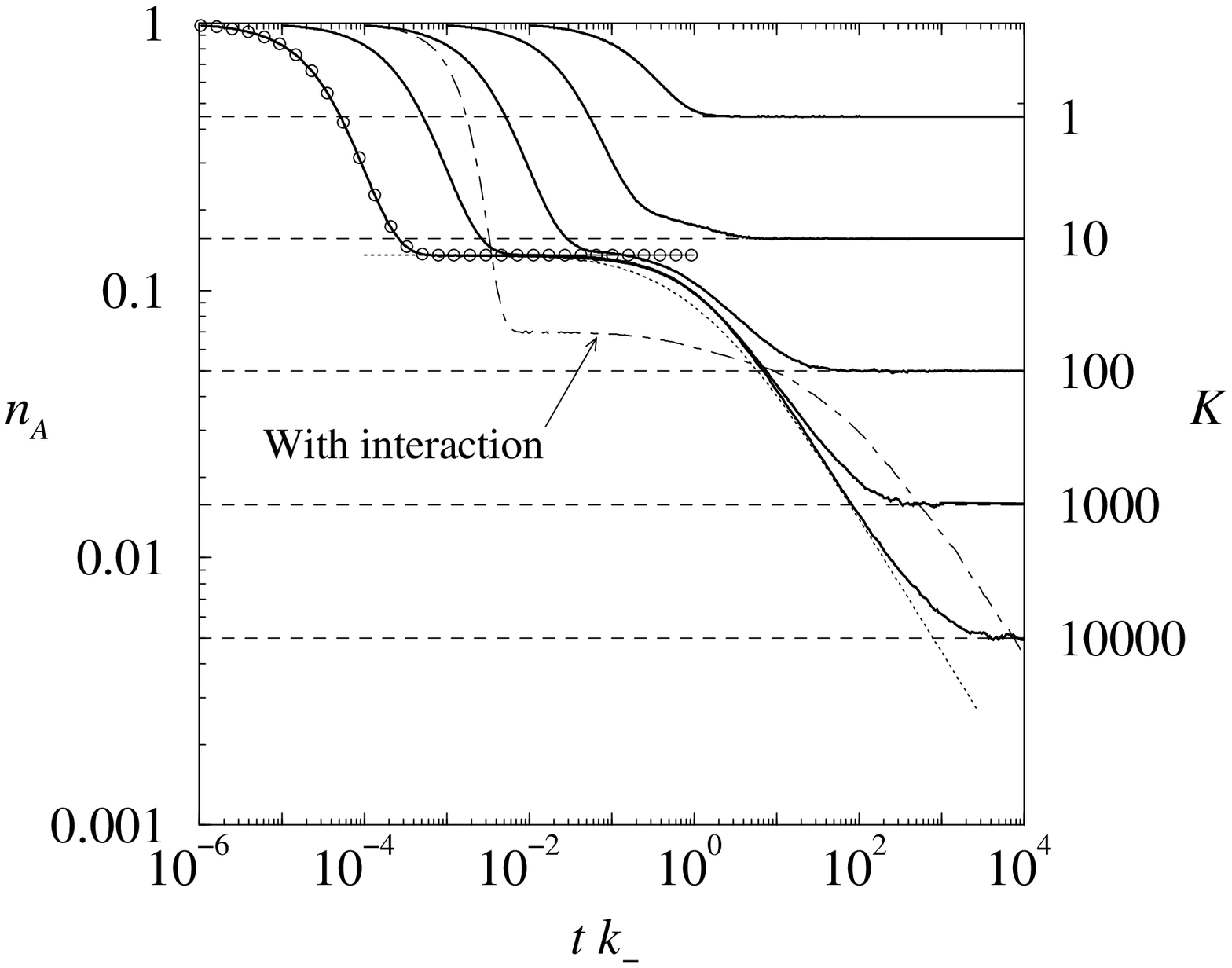}}\\
{\small Fig.~\ref{fig_relax}.}
\end{tabular}
\vspace*{0.5cm}
\caption{\label{fig_reactions}
  a) Kinesin dimers can bind one head or both heads to a one-dimensional
  lattice (tubulin protofilament). Binding sites are
  located on $\beta$-tubulin subunits (dark) while
  $\alpha$-subunits (bright) are irrelevant for our study.
  b) Reaction scheme for all binding and unbinding processes. }
\vspace*{0.5cm}
\caption{\label{fig_relax} 
  Vacancy concentration as a function of time $t$ for $K=k_+c/k_-=1,\,
  ... \, , 10000$.  Dashed lines show the steady state concentration
  $n_0=1/\sqrt{1+4K}$ at given $K$. The line with circles shows the
  short-time limit for $K=10000$ \protect\cite{mcquistan68}, and the
  dashed line the result of the reaction-diffusion model (Eq.\ 
  \protect\ref{eq_rd}). The thin dot-dashed curve shows the behaviour
  for $K=100$ with an attractive interaction between bound dimers with
  parameters $A=10$ and $B=1/10$ (a dimer is 10 times as likely to
  associate to and 10 times less likely to dissociate from a certain
  site if one of the neighbours is present) .}
\end{figure}

Motor proteins are of fundamental importance for intracellular
transport and many other biologically relevant transport processes.
Recently, a large body of data on these proteins has been collected
using a diverse set of experimental tools ranging from single-molecule
mechanics~\cite{mehta99a} to biochemical
methods~\cite{hoenger_double}. In many instances these experimental
systems are both versatile experimental techniques and interesting
non-equilibrium model systems.  One system of particular interest is a
standard method from biophysical chemistry known as ``decoration
experiments''~\cite{hoenger_double}.  Here monomeric or dimeric motor
enzymes are deposited on their corresponding molecular tracks at high
densities (see fig.~\ref{fig_reactions}a). For kinesin motors, these
tracks are microtubules, hollow cylinders usually consisting of $13$
protofilaments. The kinesin binding sites are located on the $\beta$
subunits which form a helical (wound-up rhombic) lattice with a
longitudinal periodicity of $8\,{\rm nm}$.  Decoration techniques have
traditionally been used to investigate the structure and the binding
properties of kinesin \cite{hoenger_double} {\it i.e.}\ after waiting
for the system to equilibrate the binding stoichiometry is determined
and the structure by cryo-electron microscopy followed by 3D image
reconstruction.

These experiments call for a theoretical analysis of dimer adsorption
kinetics with competing single and double bound dimers and a finite
detachment rate. For a quantitative analysis of decoration
data~\cite{vilfan-thormaehlen-frey-schwabl-mandelkow} one needs to
know the binding stoichiometry in the equilibrium state in terms of
binding constants for the first and second head of the dimer molecule.
The dynamics of the approach to equilibrium is useful to
estimate when an experimental system can be considered as
equilibrated \cite{note_estimate}.
Even more importantly, time-resolved decoration experiments (e.g.\ by
using motor enzymes labelled by some fluorescent marker) combined
with our theoretical analysis could provide new information about
reaction rates which are to date not known completely.  
Understanding the kinetics of passive motors is undoubtedly a
necessary prerequisite for studying the more complicated case of
active motors at high densities.  
The model is also interesting in its own right since it contains --~as
detailed below~-- some novel features of non-equilibrium dynamics of
dimer models which have not been discussed previously.

There seems to be convincing evidence that kinesin heads can bind on
two adjacent binding sites only in longitudinal but not in lateral
direction~\cite{hoenger_double}.  This introduces a strong uniaxial
anisotropy and distinguishes the adsorption process of protein dimers
from simple inorganic dimers. If we take into account only steric
interactions and (for now) neglect nearest neighbour attractive
interaction, we are left with a one-dimensional problem of kinesin
dimers decorating a single protofilament (one-dimensional lattice).
Then our model is defined as follows.  Kinesin is considered as a
dimeric structure with its two heads tethered together by some
flexible joint. Hence each dimer (kinesin protein) can bind one of its
two heads (motor domains) to an empty lattice site
\cite{vilfan-thormaehlen-frey-schwabl-mandelkow}.  The binding rate
$k_{+1} \, c$ for this process is proportional to the solution
concentration $c$ of the dimeric proteins. Successively, the dimer may
either dissociate from the protofilament with a rate $k_{-1}$ or also
bind its second head to an unoccupied site in front (f) of or behind
(b) the already bound head.  Since kinesin heads and microtubules are
both asymmetric structures the corresponding binding rates $k_{+2}^f$ and
$k_{+2}^b$ are in general different from each other.  The reverse
process of detaching a front or rear head occurs at rates $k_{-2}^f$
and $k_{-2}^b$. A reaction scheme with all possible processes and
their corresponding rate constants is shown in fig.\ 
\ref{fig_reactions}b.
Since there is no energy source in the system (unlike situations where
ATP is being hydrolysed), the reaction rates obey the principle of
detailed balance. As a consequence, the reaction rates are actually
not independent from each other but the ratio of the on- and off-rates
has to equal the equilibrium binding constant $K_2 = k_{+2}^b /
k_{-2}^b = k_{+2}^f / k_{-2}^f$; similarly we have $K_1 = k_{+1} /
k_{-1}$. A particular coverage of the lattice is described as a
sequence of dimers bound with both heads ($D$), one head only ($1$)
and empty sites ($0$). We denote the probabilities to find a certain
lattice site in one of these states by $2 n_D$, $n_1$ and $n_0$,
respectively.  Of course, normalisation of the probabilities requires
$n_0 + n_1 + 2 n_D=1$.

We start our analysis with the stoichiometry of the final steady
state.  Important information can be gained already by exploiting the
symmetries of the kinetic process.  The reaction scheme in fig.\
\ref{fig_reactions}b (including steric constraints) remains invariant
under the following transformation: (I) all dimers bound on a single
head are replaced by vacant sites and vice versa ($0 \leftrightarrow
1$), while dimers bound with both heads are kept as they are ($D
\leftrightarrow D$). (II) The attachment and detachment rate of the
first head are exchanged and the forward/backward binding rates of the
second head are mirrored $(k_{+1}c,k_{\pm 2}^f) \leftrightarrow
(k_{-1},k_{\pm 2}^b)$.  Since the coverage in the steady state is only
a function of $K_1 c$ and $K_2$ (see below) this symmetry implies that
the mean number of dimers attached with both heads $n_D (K_1 c, K_2)$
is invariant against interchanging the attachment and detachment rates
of the first head, $n_D (1/(K_1c),K_2) = n_D (K_1c,K_2)$.  Similarly,
the mean total number of bound heads per lattice site (often simply
called binding stoichiometry), $\nu = 2(n_1 + n_D)$, obeys the
symmetry relation $\nu(1/(K_1c),K_2)=2-\nu(K_1c,K_2)$.  Thus we
conclude that $n_D$ reaches its maximum at $K_1c=1$ where $\nu=1$.

We determine the actual value of the mean occupation numbers in the
steady state using detailed balance and the fact that the dimers have
only a hard-core interaction. Hence the probability to find a certain
sequence of $0$'s, $1$'s and $D$'s ({\it e.g.}\ ``0,1,D,D,1,D'') has to
be invariant against permutations of these states. In such a random
sequence the probabilities to find a particular state $0$, $1$ or $D$
at a certain place are given by $p_i = n_i / (n_0 + n_1 + n_D)$.
Detailed balance requires that for each pair of possible
configurations, their probabilities are in the same ratio as the
transition rates between them. Hence the ratio of probabilities to
find a sequence with a 1 or 0 at a certain place is
$p_1/p_0=k_{+1}c/k_{-1}=K_1c$. Similarly, we get for transitions between $D$ and
$01$: $p_D/p_0 p_1=k_{+2}^{b,f}/k_{-2}^{b,f}=K_2$.
These two equations, together with the normalisation condition,
uniquely determine the values $n_0$, $n_1$ and $n_D$.  The
stoichiometry, {\it i.e.}\ the total number of heads per binding site $\nu =
2(n_1+n_D)$ is given by
\begin{equation}
\label{eq8}
  \nu = 1+ \frac{K_1 c-1}{K_1 c+1}
  \left(\frac{4 K_1 K_2 c}{(1+K_1 c)^2}+1\right)^{-\frac12}\;.
\end{equation}
The number of dimers bound with both heads per lattice site reaches
its maximum $n_D^{\rm max}=(1 - 1 /\sqrt{K_2+1})/2$ for $K_1
c=1$. 

We now turn to the dynamics of the biological dimer model. In order to
avoid unnecessary complications we restrict ourselves to the limit
$K_2 \rightarrow \infty$ with $K=K_1 K_2 c$ fixed. Then our model
reduces to a dimer deposition-evaporation model where the dimers can
only bind and unbind with both heads at the same time.  The vacancy
concentration in the steady state then simplifies to
$n_0=1/\sqrt{1+4K}$.  This simplified model still captures all the
essential aspects of the non-equilibrium dynamics.  Similar dimer
adsorption models have been studied
previously~\cite{privman92,barma93,stinchcombe93}.  Privman and
Nielaba~\cite{privman92} studied the effect of diffusion on the dimer
deposition process. There are several key differences to our model,
the most important of which is that diffusion without detachment
results in a 100\,\% saturation coverage, whereas a model with
detachment leads to a limiting coverage whose value depends on the
binding constants of the first and second head (see eq.~\ref{eq8}).
This also has important implication on the dynamics as discussed
below. For example, as a consequence of a finite coverage the final
approach to equilibrium is not a power law but exponential and there
are, in addition, interesting temporal correlations in the
fluctuations in the steady state. Stinchcombe and
coworkers~\cite{barma93,stinchcombe93} studied the effect of
detachment on the adsorption kinetics but allowed for regrouping of
attached dimer molecules (two monomers that belonged to different
dimers during attachment can form a dimer and detach together).  While
such processes are allowed for some types of inorganic dimers, they
are certainly forbidden for dimer proteins like kinesin where the
linkage between its two heads is virtually unbreakable. If regrouping
is allowed the {\em steady state} auto-correlation functions for the
dimer density shows an interesting power-law decay $\propto
t^{-1/2}$~\cite{barma93,stinchcombe93}.  If it is forbidden this
power-law decay is lost (and becomes an exponential to leading order)
due to the permanent linkage between the two heads of the dimer.
Intuitively this may be understood as follows: only if regrouping of
dimers is allowed are there locally jammed configurations (N\'eel-like
states, ``101010'', with alternating occupied and unoccupied sites in
which neither attachment nor detachment of dimers is possible) in the
final steady state which slow down the dynamics.  In addition, we will
show that the autocorrelation functions of the dimer and the vacancy
occupation number show strong differences in shape and typical times
scales of relaxation.

The basic kinetic steps in the reduced model are deposition without
overlap and evaporation without regrouping of the dimers; the
effective attachment and detachment rates $k_\pm$ in terms of the
original model are $ k_{\pm} =
k_{1\pm}(k_{2\pm}^f+k_{2\pm}^b) / (k_{-1}+k_{+2}^f+k_{+2}^b) $.
Processes in which one head detaches on one side and subsequently
attaches on the other side also lead to {\em explicit} diffusion with
a rate $r_d=k_{+2}^f k_{+2}^b/(K_2(k_{-1}+k_{+2}^f+k_{+2}^b))$. This
feature is not essential since it can be incorporated in the effective
diffusion introduced later. It will hence be disregarded in what
follows (if one of the rates $k_{+2}^{f,b}$ is small, it is
negligible anyway).  Note that as a consequence of detailed balance
the diffusion of dimers is symmetric despite the asymmetry in reaction
rates.

To study the kinetics we choose the initial condition as typically
used in an experiment, namely an empty lattice.  Figure
\ref{fig_relax} shows simulation data \cite{java} for the average
vacancy concentration as a function of time for a set of binding
constants $K=k_+c/k_-\equiv K_1 K_2 c$. We find qualitatively very
different approaches to the final steady state depending on the value
of the constant $K$.  For $K \ll 1$, where the off-rate $k_-$ is much
larger than the on-rate $k_+ c$, there is no crowding on the lattice
and the dimeric nature of the molecules does not affect the approach
to equilibrium, which is, like for monomers, exponential with a decay
rate $k_-$.  In the opposite limit, $K \gg 1$, we find a {\em
two-stage relaxation} towards the steady state.  The vacancy
concentration as a function of time reveals {\em four regimes}, an
initial attachment phase, followed by an intermediate plateau, then a
power-law decay and finally an exponential approach towards
equilibrium. At short time scales, $t \ll k_-^{-1}$, when only
deposition processes are frequent but detachment processes are still
very unlikely, the kinetics of the model is equivalent to Flory's
random sequential dimer adsorption model~\cite{flory39,note2}. There
is an initial decay obeying the kinetics described in
\cite{mcquistan68,evans93} with a vacancy concentration
$n_0(t)=\exp\left(-2+2e^{-k_+c t}\right)$.  The vacancy concentration
locks at an intermediate plateau $n_0 = e^{-2}$ \cite{flory39} in the
time interval between the characteristic attachment time
$\tau_+=1/(k_+c)$ and detachment time $\tau_- = 1/k_-$.  This Flory
plateau represents a configuration in which all remaining vacancies
are isolated, causing the system to be unable to accommodate for the
deposition of additional dimers.  The secondary relaxation process
towards the final steady state is enormously slowed down.  It shows a
broad time domain with a power-law $\propto t^{-1/2}$ instead of a
simple exponential decay. Similar multi-stage relaxation processes
have been observed in dimer models with diffusion but no detachment
\cite{privman92}, the key difference being that the detachment process
implies that the steady state has a finite vacancy density and the
final approach to the steady state remains not a power law but becomes
exponential. There are also interesting similarities. In particular,
in both models a large portion of the final approach to the steady
state is mediated by the annihilation of vacancies. This behaviour can
be explained by introducing a particle representation in the following
way (analog to the adsorption-diffusion model \cite{privman92}).  We
denote each vacancy on the lattice as a ``particle'' $A$, and each
bound dimer as an inert state ($00$).  
\begin{figure}
\begin{tabular}{ll}
{\epsfxsize=0.45\textwidth \epsffile{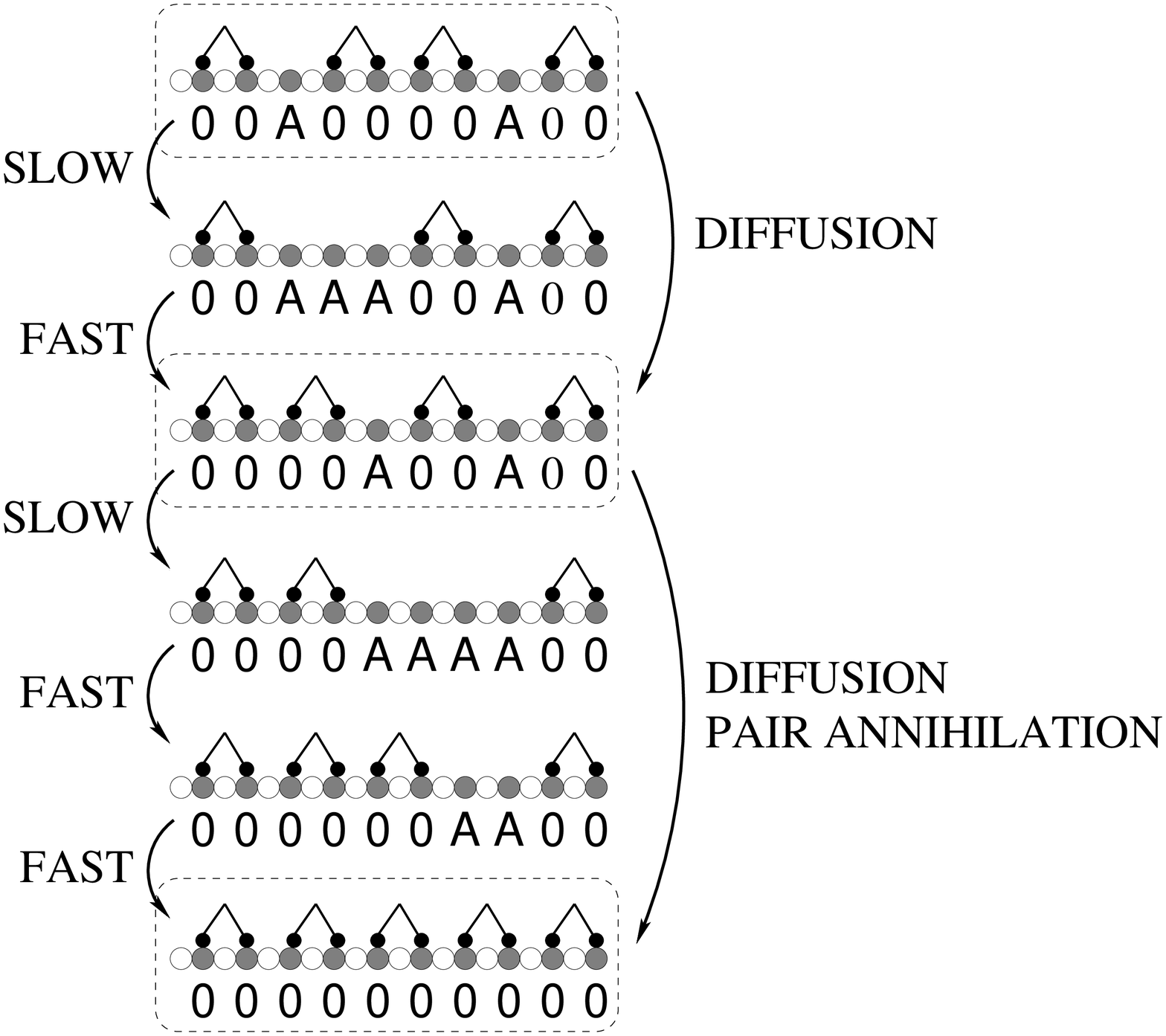}}
&
{\epsfxsize=0.48\textwidth\epsffile{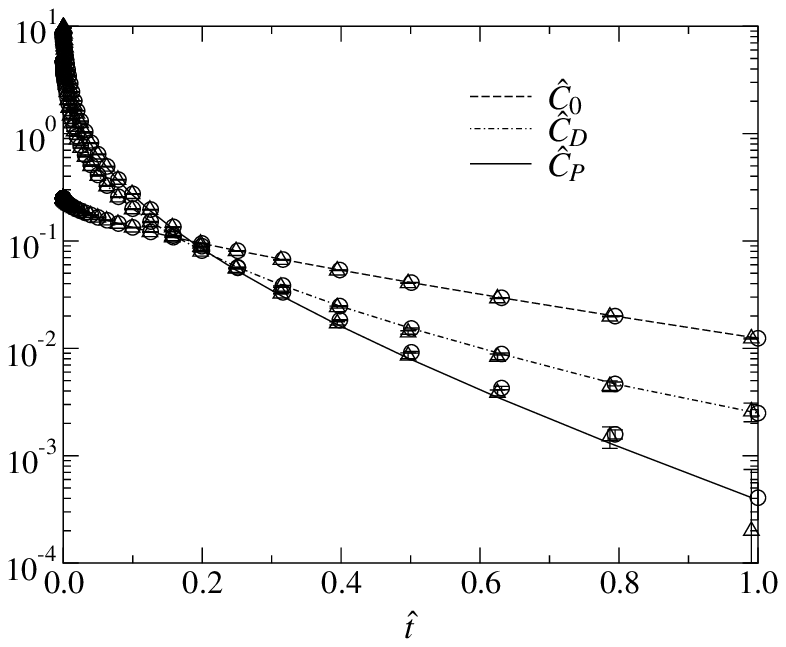}}\\
{\small Fig.~\ref{fig_cg}.}&{\small Fig.~\ref{d_fig_corr}.}
\end{tabular}
\vspace*{0.5cm}
\caption{\label{fig_cg}Time evolution of a state, consisting of 
  attachment (fast) and detachment (slow) events.  The coarse-grained
  interpretation includes only long-living states (in boxes).  The
  effective steps include diffusion, pair annihilation and pair
  creation (not shown).}
\vspace*{0.5cm}
\caption{\label{d_fig_corr}Scaled equilibrium autocorrelation functions for 
  $\hat C_0(\hat t)=K C_0(t)$, $\hat C_D(\hat t)= C_D(t)$ and  
  $\hat C_P(\hat t)=K C_P(t)$ , $\hat t=k_- t /K$. The function $\hat
  C_P(\hat t)$ is compared with the analytical calculation from
  ref.~\protect\cite{bares99_with_note} (solid line).  Simulation data were
  obtained with $K=100$ (circles, for $C_0$ and $C_D$ connected with
  dot-dashed and dashed lines) and $K=400$ (triangles).}
\end{figure}
The detachment of a dimer then
corresponds to a pair creation process $00 \to AA$, and the decoration
process to pair annihilation $AA \to 00$.  Since we consider the limit
$K \gg 1$, states with two vacancies $A$ on neighbouring sites have a
very short lifetime.  We may thus introduce a coarse-grained model by
eliminating these states (see fig.\ \ref{fig_cg}).
Then processes like $A00\to AAA\to 00A$ result in an {\em effective
  diffusion} for particle $A$ with a hopping rate $r_{\rm hop}=k_-/2$
and an effective step width of two lattice sites. {\em Pair
  annihilation}, $AA \rightarrow 00$, occurs with a rate $k_+c$. {\em
  Pair creation}, on the other hand, occurs mainly through the process
$0000\to 00AA \to AAAA \to A00A$; the corresponding rate is $k_-/3K$
per lattice site and hence largely suppressed with respect to the
annihilation process as long as the particle concentration is far
above its steady-state value.  Other processes involve more particles
and are of higher order in terms of a power series in $K^{-1}$. They
are negligible for $n_A \ll 1$ and $K\gg 1$. In summary, for $K \gg
1$, our dimer model can be mapped onto a one-particle
reaction-diffusion model $A+A \rightarrow 0$. Pair creation processes,
$0 \rightarrow A+A$, are highly suppressed and do not play any role
until the system comes close the steady state.

Models like the simple reaction-diffusion model $A+A \rightarrow 0$
show interesting non-equilibrium
dynamics~\cite{privman97,mattis98,schuetz98}.  One can
show~\cite{torney83,toussaint83} that asymptotically the particle
density $n_A (t)$ decays algebraically, $n_A (t) \propto t^{-1/2}$
which nicely explains the slow decay observed in simulation data (see
fig.\ \ref{fig_relax}).  Note that a mean-field like rate equation
approach would predict $n_A (t) \propto t^{-1}$. In our analysis we
can even go beyond the asymptotic scaling analysis and try to compare
with exact solutions of the model for a random initial distribution
with density $p$ by Krebs {\it et al.}~\cite{krebs95}.  They find
(adapted to our situation with two-site hopping)
\begin{eqnarray}
\label{eq_rd}
  n_A (t) = \frac{1}{2\pi}
  \int\limits_0^2 du \, 
  \frac{ \sqrt{u(2-u)} \, p^2 \, e^{-16u r_{\rm hop} t}}
       {u \, \left( u (\frac12-p) + p^2 \right)} \;.
\end{eqnarray}
Its asymptotic limit (first determined by Torney and McConnel~\cite{torney83})
reads $n_A=(32\pi r_{\rm hop} t)^{-1/2}$; note that it is
independent of the initial particle concentration in the Flory plateau $p$.

Our Monte-Carlo data (see fig.\ \ref{fig_relax}) are in excellent
agreement with the predictions of eq.\ \ref{eq_rd}. Minor deviations
at times between the plateau and the power-law decay are due to the
assumption of a random particle distribution underlying the derivation
of eq.\ \ref{eq_rd}; the random sequential adsorption process leads to
some particle correlations in the intermediate plateau regime
\cite{evans93}, but they do not affect the asymptotic behaviour as they
are only short-ranged.  The results from the $A+A \rightarrow 0$ model
also become invalid for very long times where the particle
concentration comes close to its equilibrium value. In this limit the
dynamics becomes scale-invariant \cite{racz85}. The particle
concentration can be written in scaling form $n_A(t)=K^{-1/2} \hat
n(r_{\rm hop} t /K)$ with a diverging characteristic time scale
$\tau_K \propto K$.

A limitation of the above mapping becomes evident if one considers the
equilibrium autocorrelation functions.  Contrary to conventional
models there are three different autocorrelation functions with
different functional forms (fig.~\ref{d_fig_corr}).  $C_0(t)=\left<
\hat n_{0_{i}} (t_0) \hat n_{0 _{i}} (t_0+t) \right> - \left< \hat
n_{0_{i}} \right>^2$ describes the correlation function of the
probability to find a vacancy at a certain lattice site.  $C_D$ is the
equivalent quantity defined from the probability to find a dimer on a
certain pair of sites $\hat n_{D_{i,i+1}}$ and $C_P$ from the
probability to find a vacancy on at least one from a pair of
neighbouring sites, $\hat n_{0_{i}}+\hat n_{0_{i+1}}-\hat
n_{0_{i}}\hat n_{0_{i+1}}$.  For $K\gg 1$ the autocorrelation
functions become scale invariant as well.  Their scaling form reads
$\hat C_0(\hat t)=K C_0(t)$, $\hat C_D(\hat t)= C_D(t)$ and $\hat
C_P(\hat t)=K C_P(t)$ with $\hat t=k_- t /K$. The latter corresponds
to the autocorrelation function in a reaction-diffusion model
$A+A\leftrightarrow 0$, which has recently been calculated
analytically in ref.~\cite{bares99_with_note}
\begin{equation}
\label{d_eq_cpscale}
\hat C_P(\hat t)=\left( \frac {e^{-2 \hat t}}{\sqrt {2 \pi \hat t}} -
  \mbox{Erfc}\,\sqrt{2 \hat t} \right)  \mbox{Erfc}\,\sqrt{2 \hat t}\;, \qquad
  {\rm with}\quad \mbox{Erfc}\,(x)=\frac 2 {\sqrt{\pi}} \int_x^\infty
  e^{-y^2} dy\;.
\end{equation}
The other two functions decay on the same time-scale, but with
different prefactors. The reason is that even if a pair of vacancies
annihilates, the system still keeps memory on whether the surrounding
dimers were located on even or odd locations and this gives those
correlation functions that distinguish between even and sites a longer
decay time.

Finally, we would like to note that an attractive interaction between
attached dimers plays an important role in some cases.  In the case of
kinesin, there has been an observation of coexisting empty and
decorated domains which can only be explained by an attractive
interaction \cite{vilfan-thormaehlen-frey-schwabl-mandelkow}.  Similar
observation has been done on actin decorated with myosin
\cite{orlova97} and tropomyosin \cite{wegner79}.  We introduce the
interaction by assuming that a dimer is more likely to bind to a pair
sites if one or two neighbours are already bound.  The binding rate
then becomes $A k_+$ (one neighbour bound) or $A^2 k_+$ (both
neighbours bound).  Similar, we assume that a dimer with one bound
neighbour dissociates with rate $B k_-$ and that a dimer with two
bound neighbours dissociates with rate $B^2 k_-$.  This interaction
changes both relaxation stages quantitatively.  First, the vacancy
concentration on the intermediate plateau lowers since the interaction
improves the formation of contiguous clusters during the first stage.
Second, the diffusional relaxation slows down since the detachment
rate decreases.  And finally, the equilibrium vacancy concentration
decreases.  Nevertheless, interacting models show the same two-stage
relaxation behaviour.  An example of a model with interaction is shown
by the dot-dashed line in fig.~\ref{fig_relax}. We therefore expect
that many qualitative conclusions from this article will apply to a
much wider range of biological adsorption problems such as the binding
of double-headed myosin \cite{orlova97} or tropomyosin
\cite{wegner79}. 

We would like to thank E. Mandelkow and A. Hoenger for helpful discussion about
microtubule decoration experiments and their relevance for studying
motor proteins. We have also benefited from discussions with T.
Franosch, J. Santos, G.  Sch{\"u}tz and U.C. T{\"a}uber. Our work has
been supported by the DFG under contract nos.\ SFB~413 and FR~850/4-1.
E.F. acknowledges support by a Heisenberg fellowship from the DFG
under Grant no.~FR~850/3-1.


\begin{thebibliography}{10}
\bibitem{note_estimate}
With reaction rates from \cite{hancock99} ($k_-\approx 0.001\,{\rm s}^{-1}$
  with AMP-PNP or without nucleotide) and assuming that there is no explicit
  diffusion ($r_d \ll k_-$) the relaxation time for a high kinesin
  concentration ($K=100$) can be as long as $10^5\,{\rm s}$. On the other hand,
  in the presence of ADP ($k_-\approx 1\,{\rm s}^{-1}$) and $K=10$ the
  relaxation time can be as short as $10\,{\rm s}$. Some care has to be taken
  as these results were obtained with kinesin attached to beads.
\bibitem{note2}
The case without explicit diffusion corresponds to the
  ``standard''~\cite{flory39} dimer filling model and the diffusive case to the
  ``head-on'' dimer filling model~\cite{page59,nord90} with the final vacancy
  concentration $0.1233$. The final configuration is independent of the
  asymmetry in the binding rates.
\end{thebibliography}
\end{document}